\documentclass[twocolumn,showpacs,superscriptaddress,pra,floatfix,nofootinbib]{revtex4-1}
\usepackage{amsmath}
\usepackage{amssymb}
\usepackage{latexsym}
\usepackage{array}
\usepackage{amsfonts}
\usepackage{mathrsfs}
\usepackage{color}
\usepackage{hyperref}
\usepackage{graphicx}
\usepackage{multirow}
\def\be{\begin{equation}}
\def\ee{\end{equation}}
\def\bea{\begin{eqnarray}}
\def\eea{\end{eqnarray}}
\def\ben{\begin{equation*}}
\def\een{\end{equation*}}
\def\bean{\begin{eqnarray*}}
\def\eean{\end{eqnarray*}}
\def\bma{\begin{mathletters}}
\def\ema{\end{mathletters}}
\def\bi{\begin{itemize}}
\def\ei{\end{itemize}}

\newcommand{\ket}[1]{ | \, #1 \rangle}
\newcommand{\bra}[1]{ \langle #1 \, |}
\newcommand{\proj}[1]{\ket{#1}\bra{#1}}

\begin{document}
\title{Quantum Byzantine Agreement via Hardy correlations and entanglement swapping}
\author{Ramij Rahaman}\email{ramijrahaman@gmail.com}
\affiliation{Department of Mathematics, University of Allahabad, Allahabad 211002, U.P., India}
\author{Marcin Wie\'sniak}\affiliation{Institute of Theoretical Physics and Astrophysics, University of Gda\'nsk,\\ PL-80-952 Gda\'nsk, Poland}
\author{Marek \.Zukowski}\affiliation{Institute of Theoretical Physics and Astrophysics, University of Gda\'nsk,\\ PL-80-952 Gda\'nsk, Poland}


\begin{abstract}
We present a device-independent quantum scheme for the {\em Byzantine Generals} problem. The protocol is for three parties. Party $C$ is to send two identical one bit messages to parties $A$ and $B$. The receivers $A$ and $B$ may exchange two one bit messages informing the other party on the message received from $C$. A bit flipping error in one of the transmissions, does not allow the receiving parties to establish what was the message of $C$. Our quantum scheme has the feature that if the messages of the Byzantine protocol are readable (that is give an unambiguous bit value for any of the receivers), then any error by $C$ (cheating by one of the commanding general) is impossible. $A$ and $B$ do not have to exchange protocol messages to be sure of this.
\end{abstract}

\maketitle
Quantum Mechanics (QM) allows to encode and process information in ways inaccessible to classical theories with non-contextual hidden variables. One such strictly quantum resource is entanglement, a phenomenon in which individual parts of a quantum system can be described only in reference the other ones. There are multiple consequences of entanglement. The fundamental one is the exclusion of theories with local hidden variables (LHVTs, where non-contextuality is justified by spatial separation). The practical ones are advantages in various communication tasks, as compared to classical protocols. Examples of such advantage include security of cryptographic key distribution or secret sharing, or communication complexity reduction in a distributed computing system.

A procedure, in which quantum correlations can also be useful, deal with faults of components of a vast computing network. A reliable distributed computing system must be able to cope with a failure of some of its components. A failing component can behave arbitrarily and may send conflicting information to different parts of the computing system.

The abstract formulation of the problem is put in a form of generals of the Byzantine Army communicating with each other. The commanding general $C$ (say, Clausevitz) first sends a message $(m_C)$ to $A$ (for Alexander) and $B$ (for Buonaparte) whether to attack or retreat. Next, $A$ and $B$ exchange the messages to confirm what they received. The generals must reach a consensus among themselves based on the messages exchanged \cite{PSL80,LSP82}. The problem is complicated by the fact that one (and only one) of the players can be traitor. He may try to convince the loyal general(s) for malevolent action by sending corrupt message. The solution to the problem must allow (i) all the loyal generals to agree upon a common plan of action. Also, (ii) if the commanding general is loyal then all the loyal generals must obey the order he sends. In more mundane terms Byzantine Agreement (BA) is about information transmission. An electronic device $C$ should send two identical bits to devices $A$ and $B$, but there might be a bit-flipping error (which causes the messages to be different). $A$ gets a confirmation bit from $B$ which is to inform what was the value received by $B$ from $C$. However also in this transfer there may be a bit flipping error in the transmission form $A$ to $B$. Similarly $B$ gets from $A$ a bit which is supposed to be the one received by $A$ form $C$, but there might be an error in the transmission. How can A know that the possible discrepancy of the messages from $B$ and $C$, which $A$ received, is due to an error by $B$ or by $C$? The same applies to $B$, which device sent a bit-flipped message $C$ or $A$ (that is, which device is faulty)? However, $C$ is the controlling unit, therefore $A$ must act according to $C$'s message, if its message was identical to $A$ and $B$ (a correct operation of $C$). The same applies to $B$. The assumption of just one error is motivated by the fact that double errors are rare.

The BA problem is about faulty transmission, not failure of transmission. Therefore, any attempt of jamming/cutting the communication link for any players leads to a different problem other than BA. In other words, prevention against jamming/cutting does not lie in the scope of this problem. One should also notice that the BA problem is not a cryptographic problem and in the standard case at most one traitor is assumed.

It was shown that an unconditionally secure scheme for the BA problem is probably unsolvable by means of classical resources only \cite{PSL80,LSP82,FLM86,FGMO01}. However, a partial quantum solution was suggested in Ref. \cite{FGM01,Cab02,Cab03a,Cab03,IG04,GBKCW08,BCZ10} in which the condition (ii) had to be compromised to (ii'): all loyal generals either follow the same plan, or abort any action. This modification of the original problem is known as detectable Byzantine agreement (DBA) or detectable broadcast \cite{FGM01,GBKCW08}. Here, we present a secure protocol for the original BA problem. Our protocol is based on Hardy's paradox \cite{Har92}, which disproves the possibility of having a LHVTs description of quantum correlations like Bell \cite{Bel64}, but without inequalities.

Let us begin with recalling the original Hardy `paradox' \cite{Har92}. Consider a bipartite system and a choice of two local observables $U_x$ and $D_x$, where $(x=1,2)$ denotes the systems, with outcomes $\pm 1$. Let $P(y_1,y_2|Y_1,Y_2)$ denote the joint probability that the measurements $(Y_1, Y_2)$ gave the results $(y_1,y_2)$. Hardy noticed that the following four conditions can be satisfied, for some $q$ and some quantum states:
\bea\label{hardy0}
&P(-1,-1|D_1,D_2)=0.&\nonumber\\
&P(+1,+1|D_1,U_2)=0,&\nonumber\\
&P(+1,+1|U_1,D_2)=0,&\nonumber\\
&P(+1,+1|U_1,U_2)=q>0,&\\ \nonumber
\eea
However, this set of conditions cannot be satisfied in LHVTs, and therefore by any separable state. In these theories, the last condition of (\ref{hardy0}) says that each of the subsystems can yield result ``$+1$'' under measurement $U$. In such a case, the second and the third condition of (\ref{hardy0}) tell us that the subsystems will yield ``$-1$'' under $D$, which is in contradiction with the first condition. However, these conditions can be met by entangled states \cite{Har92}.

Let us find state $\rho$, for which the conditions for Hardy-type argument given in (\ref{hardy0}) are satisfied for given two pairs of observables $(U_k, D_k)$, $k = 1, 2$.
Denote by $\ket{x}$ and $\ket{x^{\perp}}$ eigenstates of a Pauli-type observable $X$ with eigenvalues $+1$ and $-1$, respectively.
Using such notation, any state, which satisfies conditions (\ref{hardy0}), has to be orthogonal to the following three product states $\ket{\phi_0}=\ket{d_1^{\perp}}\ket{d_2^{\perp}}$, $\ket{\phi_1}=\ket{u_1}\ket{d_2}$ and $\ket{\phi_2}=\ket{d_1}\ket{u_2}$ associated with the three zero probabilities of (\ref{hardy0}) and is non-orthogonal to $\ket{\phi_3}=\ket{u_1}\ket{u_2}$ associated with the non-zero probability of (\ref{hardy0}).

$\ket{\phi_0}, \ket{\phi_1}$ and $\ket{\phi_2}$ span a three-dimensional subspace $\mathcal{S}$. Therefore, to satisfy the conditions (\ref{hardy0}), a state has to be confined to an one dimensional subspace $\mathcal{S}^{\perp}$ of $\mathcal{C}^2\otimes \mathcal{C}^2$, which is orthogonal to $\mathcal{S}$. Therefore, $\rho$ must be a unique, pure and entangled \cite{Kar97}. We shall denote it as $\ket{\psi^H}$. The four product states $\{\ket{\phi_i}\}_{i=0}^3$ are linearly independent, hence by Gram-Schmidt orthogonalization procedure one can find a basis $\{\ket{\phi'_i}\}_{i=0}^3$, in which the Hardy state $\ket{\psi^H}=\ket{\phi'_3}$ is its last member:
\be
\ket{\phi'_0}=\ket{\phi_0},\ket{\phi'_i}= \frac{\ket{\phi_{i}}-\sum^{i-1}_{j=0}\langle \phi'_j|\phi_{i}\rangle\ket{\phi'_j}}
{\sqrt{1-\sum^{i-1}_{j=0}|\langle \phi'_j|\phi_{i}\rangle|^2}}, i=1,2,3.\label{hardys2}
\ee
As $D_j\neq U_j$, one must have
\be\label{obser2}
\ket{d_j}=\alpha_j\ket{u_j}+\beta_j \ket{u_j^{\perp}},\ket{d_j^{\perp}}=\beta^*_j\ket{u_j}-\alpha^*_j \ket{u_j^{\perp}},
\ee with $|\alpha_j|^2+|\beta_j|^2=1$ and $0<|\alpha_j|<1$, for $j=1,2$. Thus, the probability
 $q$ in the conditions (\ref{hardy0}) reads
 \ben\label{value_q}
q = |\langle \psi|\phi_3\rangle|^2 = 1-\sum_{i=0}^2|\langle \phi'_i|\phi_3\rangle|^2=
 \frac{|\alpha_1\alpha_2|^2|\beta_1\beta_2|^2}{1-|\alpha_1\alpha_2|^2}.
\een
Its maximum possible value is $\frac{5\sqrt{5}-11}{2}$ for $|\alpha_1|=|\alpha_2|=\sqrt{\frac{\sqrt{5}-1}{2}}$ \cite{Jor94}.

Let us define the unique Hardy state by $\ket{\psi^*}$ for which $q$ achieves its maximum value. A recent result by Rabelo {\em et.al.} \cite{RZS12}, tells us that, for $q_{max}=\frac{5\sqrt{5}-11}{2}$ the state of of any systems is equivalent to $\ket{\psi^*}_{12}\otimes\ket{\eta}_{1'2'},$ where $\ket{\eta}_{1'2'}$ is an arbitrary bipartite junk state for some other systems. State $\ket{\psi^*}_{12}$ is unique, and any expansion of the Hilbert spaces of local systems leads to a factorisable extension.

Therefore, Hardy conditions with $q_{max}$ constitute a device independent test uniquely pinpointing $\ket{\psi^*}_{12}$ as responsible for the correlations. Note that the value of $q$ is determined by the choice of local observables used to define the Hardy conditions, and so is the state.

{\em Protocol for Byzantine Agreement:} Let the commanding general $C$ send a one bit message $m_C$ to two generals, $A$ and $B$. Let us denote $m_{CA}$ and $m_{CB}$ the bits received by $A$ and $B$, respectively (they will be the BA protocol bits, all other information exchange to transmit these is treated as auxiliary). After receiving the message bit $(m_{CG})$, the general $G$, where $G=A,B$, sends bit $m_{GR}$ ($G\neq R=A,B$) to the other general $R$ to inform about the message ($m_{CG}$) he received from $C$. In a three-party BA problem at most one player can be traitor. So, if $m_{AB}=m_{BA}$, then all three players are {\em loyal} and there is no problem in Byzantine agreement. But if, $m_{AB}\neq m_{BA}$, then one of the {\em general} (from generals $A, B$ and $C$) must be a traitor. The main goal of this problem is to find out the traitor. In order to accomplish this, we propose a quantum scheme which based on some novel features of Hardy paradox. One should note that there is an asymmetry in the problem. $C$ is the commanding general, he just sends the messages and not receives any (protocol) messages from $A$ and $B$. Whereas, the roles of $A$ and $B$ are symmetric. They receive as well as send messages. Also, both $A$ and $B$ (at least the loyal one) are interested to know who is the traitor in contrast, $C$ is not. Cheating by $C$ is sending two different messages to $A$ and $B$. Whereas, cheating by $A$ is sending to $B$ a message which is opposite to the one he received from $C$. Like $A$, we have a symmetric definition of cheating by $B$. All {\em classical} `protocol messages' are only {\em sent} by $C$, while $A$ and $B$ exchange confirming messages between themselves only. There is some non protocol, auxiliary, information which may be sent to $C$, to test and operate the quantum links only. What is important we assume that $C$ sends the classical messages to $A$ and $B$ on his decision retreat/attacks using only quantum communication methods. As we shall see, this does not need to be the case for the message exchange between $A$ and $B$.

Our quantum scheme for BA problem consists with two symmetric sub-protocols one for $A$ and another for $B$. Both $A$ and $B$ prepare as well as distribute the necessary resources for their part of the protocol. Due to the symmetry, here we describe only $A$'s part of the protocol in detail.

Let us begin with the part of the protocol which is done on qubits distributed by $A$. The goals of $A$'s protocol are:
\begin{enumerate}
 \item to allow $C$ to send the message $m_{CB}$ to $B$,
 \item to allow $A$ to check what was the message $C$ sent to $B$.
\end{enumerate}
\begin{description}
 \item[S1] {\em Distribution of resources:} For simplicity, consider $U_A=U_B=U_C=U$ and $D_A=D_B=D_C=D$ as the protocol settings. In the bases related with the $U$-measurements, the Hardy state given in Eq. (\ref{hardys2}), can be written as,
\be\begin{split}\label{hardystate}
\ket{\psi^H}&=x_{00}\ket{u}_1\ket{u}_2
+x_{01}(\ket{u}_1\ket{u^{\perp}}_2\\&+\ket{u^{\perp}}_1\ket{u}_2)+x_{11}\ket{u^{\perp}}_1\ket{u^{\perp}}_2,
\end{split}\ee
where, $x_{00}=\frac{|\alpha\beta|^2}{\sqrt{1-|\alpha|^4}}$, $x_{01}=-\frac{\alpha^*\beta|\alpha|^2}{\sqrt{1-|\alpha|^4}}$, and $x_{11}=-\frac{{\alpha^*}^2\beta^2\sqrt{1-|\alpha|^4}}{|\alpha\beta|^2}$, with $\alpha_1=\alpha_2=\alpha$ and $\beta_1=\beta_2=\beta$.

Initially, $A$ shares a large number of copies of two-qubit maximally entangled states $\ket{\Phi^+}=\frac{1}{\sqrt{2}}[\ket{uu}+\ket{u^{\perp}u^{\perp}}]$, say, $6N$ with both $B$ and $C$. Each general keeps the record of his qubits, by writing down detection times, settings and results.

\begin{description}
 \item[a] {\em Conversion of $\ket{\Phi^+}$ to $\ket{\psi^H}$:} $A$ randomly selects $4N$ copies of $\ket{\Phi^+}$, $2N$ shared with $B$ and $2N$ shared with $C$ and prepares one ancilla qubit $\ket{u}_a$ for each selected copy. Next, $A$ applies a two-qubit unitary operation $\mathbb{U}$ on each pair (an ancilla and his system qubit, for a randomly selected copy of $\ket{\Phi^+}$).
 \be
(\mathbb{U}^{a1}\otimes \mathbb{I}^2)\ket{u}_a\ket{\Phi^+}_{12}=\frac{1}{\sqrt{2}}\left[\ket{u}_a\ket{\psi^H}_{12}+ \ket{u^{\perp}}_a\ket{\psi'}_{12}\right],
\ee
where \bean\mathbb{U}\ket{uu}=x_{00}\ket{uu}+x_{01}\ket{uu^{\perp}}+x_{01}^*\ket{u^{\perp}u}+x_{11}^*\ket{u^{\perp}u^{\perp}}, \\ \mathbb{U}\ket{uu^{\perp}}=x_{01}\ket{uu}+x_{11}\ket{uu^{\perp}}-x_{00}^*\ket{u^{\perp}u}-x_{01}^*\ket{u^{\perp}u^{\perp}}\eean and \ben \ket{\psi'}=x_{01}^*\ket{uu}-x_{00}^*\ket{uu^{\perp}}+x_{11}^*\ket{u^{\perp}u}-x_{01}^*\ket{u^{\perp}u^{\perp}}.\een
After this, $A$ measures each ancilla in basis $\{\proj{u},\proj{u^{\perp}}\}$ and discards the runs \footnote{each run associated with each copy of the entangled state.} (say $R_1$) with measurement outcome $\ket{u^{\perp}}$. If the outcome is $\ket{u}$ (which can happen with probability $\frac{1}{2}$), the corresponding maximally entangled state shared between $A$ and $B$, or $C$, collapses to a shared Hardy state $\ket{\psi^H}$.

\item[b] {\em Creation of Hardy state $\ket{\psi^H}$ between $B$ and $C$:} From the remaining $8N$ copies of $\ket{\Phi^+}$, $A$ randomly selects and prepares $4N$ pairs of copies $\{\ket{\Phi^+}_{12},\ket{\Phi^+}_{34}\}$, such that in each pair the first $\ket{\Phi^+}_{12}$ is shared between $A$ and $B$ and the second $\ket{\Phi^+}_{34}$ between $A$ and $C$. Qubits `$1$' and `$3$' are in $A$'s hand for each such pair of copies. $A$ applies a two outcome joint measurement $\{M,I\smallsetminus M\}$ on her two qubits for each such selected pair of copies $\{\ket{\Phi^+}_{12},\ket{\Phi^+}_{34}\}$. Here, $M=|{\psi^H}^*\rangle\langle {\psi^H}^*|$ with \ben
\ket{{\psi^H}^*}=x_{00}^*\ket{u u}+x_{01}^*\left(\ket{uu^{\perp}}+\ket{u^{\perp}u}\right)+x_{11}^*\ket{u^{\perp}u^{\perp}}.
\een
In the measurement, if $M$ clicks (which can happen with probability $\frac{1}{4}$), the associated pair of maximally entangled states $\{\ket{\Phi^+}_{12},\ket{\Phi^+}_{34}\}$ collapses to a Hardy state $\ket{\psi^H}$ shared between $B$ and $C$:
\ben(M^{13}\otimes \mathbb{I}^{24})\ket{\Phi^+}_{12}\ket{\Phi^+}_{34}=\frac{1}{2}|{\psi^H}^*\rangle_{13}\otimes|{\psi^H}\rangle_{24}.\een
$A$ discards those runs (say, $R_2$) where measurement outcome was not $M$.
\end{description}
The runs $R_3=\{R_1\}\bigcup R_2$ are discarded. and they are totally useless for rest of the protocol. This list is distributed among the generals. Each of the party receives approximately $2N$ ($2N^{\approx}$, for short-hand) qubits and consequently each pair of parties shared $N^{\approx}$ copies of $\ket{\psi^H}$ between them. Neither $B$ nor $C$ can know which of his qubits (from their own $2N^{\approx}$-qubits) were entangled with $A$'s qubits and which of them were entangled with the other one of the pair, while $A$ has full knowledge about the correlation links of all $3N^{\approx}$ pairs of qubits {\em i.e.}, all $3N^{\approx}$ copies of $\ket{\psi^H}$.
\item[S2] {\em The actual actions on qubits distributed by $A$:}
As said earlier $A$ makes measurements in random $U$ and $D$ bases (test settings) on all of her $2N^{\approx}$-qubits form shared Hardy states (with $B$ and $C$), so does $B$. Whereas $C$, if he wants the protocol to run is to make, say, $75\%$ of their randomly chosen measurements in the message basis (in a prearranged manner), and the rest in random bases (test settings). $C$ chooses basis $U$ for the message `$m=0/${\em Yes}' or $D$ for `$m=1/${\em No}'. They announce their {\em results} (not settings) of {\em all} runs, immediately after each measurement. Let $L$ and $L_1$ be the lists of runs where $C$ have chosen the measurements in {\em message basis} and {\em random bases} accordingly. Before going to next step $C$ sends the list $L/L_1$ to both $A$ and $B$.
\item[Note]
An important requirement of the protocol is that generals $B$ and $C$ make their protocol measurements immediately after they receive qubits. This can be forced by requesting $A,B$ and $C$ to announce the results, not settings, immediately, for each run in a random sequence, so the each partner has probability $1/3$ to be first to announce. General $A$ does his protocol measurements after his preparation measurements (swappings, projections), in a sequential manner. In the case of entanglement swapping run $A$ does not have to make measurements of his leftover qubits, but should nevertheless announce some `{\em phony}' results, say random. All this is to preclude delayed choice operation of any of the partners, and to hide, at results revealing stage, who was connected with whom. Additionally, there is no need to ``{\em store}" the qubits. This makes the protocol more feasible.

\item[S3] {\em Convey classical message through measurement settings:} After receiving the list $L/L_1$ and measurement results from $C$, $A$ can easily find out the message basis of $C$. Hence, he can read the messages $m_{CB}$. $A$ checks for what choice of uniform measurement setting $(U/D)$ on $C$'s side the measurement data of all the correlated qubit pairs between him and $C$ from $L$ has no contradiction with Hardy's conditions (\ref{hardy0}). If $C$ sends consistent data there must be one uniform setting for $L$ and that setting represents the message $m_{CB}$.

 Next $A$ reveals to $B$ and $C$ which runs were connecting whom {\em i.e.}, the {correlation links}\footnote{Position and parties information of each pair of correlated qubits associated to a Hardy state shared between X and Y.} for all Hardy states. $A$ cannot cheat in this because this would lead to no transmission of $m_{CB}$. However, $C$ does not know before making measurements with whom he was connected. So, he has no option other than choosing the same message settings all the time, if he wants to send a consistent (readable) message to B. Otherwise communication of the BA protocol bit $m_{CB}$ fails. Recall, that $A$ reveals the links, only after measurement results are announced by all parties.

\item[S4] Upon knowing the correlation links with $C$'s qubits, $B$ can easily read the message $m_{CB}$ by verifying Hardy's conditions for the correlated qubits pairs he shared with $C$ from list $L$. Like $A$, $B$ also checks for what choice of uniform setting between $U$ and $D$ on $C$'s side the measurement data for all these correlated qubits from $L$ satisfy the Hardy's conditions \ref{hardy0}. If the links revealed by $A$ are genuine and $C$ are honest then the measurement data will be consistent for one of the choices of a uniform setting (say, $V\in\{U,D\}$). Hence, he can read the message $m_{CB}=V$. $B$ and $C$ can check whether they were indeed sharing Hardy states with $A$ and themselves (also $A$ sent them the correct correlation links) by exchanging their measurement results and settings, as part of the settings are test settings. They can also ask $A$ to send his settings and results to test whether the Hardy's correlations they shared with $A$ are genuine or not. 
 If the shared Hardy states are all genuine then the measurement data for each pair of correlated qubits agree with Hardy's conditions (\ref{hardy0}).
\item[S5] {\em The actual actions with qubits distributed by $B$:} All is symmetric with respect to the above (only the roles of $A$ and $B$ are exchanged and the BA protocol message in question is now $m_{CA}$). If transmission goes well, $A$ and $B$ know the message $m_{CA}$, which was meant only for $A$.
\end{description}
{\em Betrayal consequences:} If $C$ is a traitor, the other partners, who are by definition now loyal, know this immediately, they even do not have to exchange any messages anymore. If $A$ is a traitor, he may try to fool $B$ by sending him a classical message $m_{AB}$ which is opposite to $m_{CA}$. But this is useless, because $B$ knows $m_{CA}$ from the quantum protocol. It may also happen that the messages (readable) $m_{CB}$ and $m_{CA}$ $B$ received are not same {\em i.e.}, $m_{CB}\neq m_{CA}$. This is possible only if {\em $A$ filliped $C$'s preparation basis} ($U_C\leftrightarrows D_C$) at the time of preparation of the Hardy states between $B$ and $C$ in step S1(a). That is, instead of genuine Hardy states (\ref{hardystate}) $A$ generates copies of the following correlation between $B$ and $C$.
 \be\begin{split}\label{fake}
\ket{\chi}&=x_{00}\ket{u d}
+x_{01}(\ket{u d^{\perp}}+\ket{u^{\perp}d})+x_{11}\ket{u^{\perp}d^{\perp}}.
\end{split}\ee
To this end, $A$ employs the measurement $\{M',\mathbb{I}\smallsetminus M'\}$ instead of $\{M,\mathbb{I}\smallsetminus M\}$. Here, $M'=\proj{\chi^*}$ with \ben\begin{split}\ket{\chi^*}=&x_{00}^*\ket{u}\left(\alpha^*\ket{u}+\beta^*\ket{u^{\perp}}\right)\\&+
x_{01}^*\left[\ket{u}\left(\beta\ket{u}-\alpha\ket{u^{\perp}}\right)+\ket{u^{\perp}}\left(\alpha^*\ket{u}+\beta^*\ket{u^{\perp}}\right)\right]
\\&+x_{11}^*\ket{u^{\perp}}\left(\beta\ket{u}-\alpha\ket{u^{\perp}}\right).\end{split}\een
In the measurement, if $M'$ clicks then the corresponding pair of maximally entangled states $\{\ket{\Phi^+}_{12},\ket{\Phi^+}_{34}\}$ collapses to a copy of the state $\ket{\chi}$ shared between $B$ and $C$ (for detail see step S1(a)):
\ben(M'^{13}\otimes \mathbb{I}^{24})\ket{\Phi^+}_{12}\ket{\Phi^+}_{34}=\frac{1}{2}|{\chi}^*\rangle_{13}\otimes|{\chi}\rangle_{24}.\een
Due to the symmetry of the correlation of $\ket{\chi}$ general  $B$ reads out a filliped message $m'_{CB}$ of an opposite bit-value with respect to what $C$ wanted to convey him. But the correlations characteristic for $\ket{\chi}$ cannot reproduce all the conditions of (\ref{hardy0}). So, by verifying Hardy's condition with $C$ and $m'_{CB}\neq m_{CA}$ partner $B$ can easily find out that $A$ is the traitor. Note that, since $C$ does not have any prior knowledge of the correlation links before his measurements, so he cannot selectively choose different measurements and convey two different messages to $A$ and $B$.

The same holds in the case if $B$ is a traitor. Thus the traitor has no chance to hide.

A moment of thought allows one to find out that the protocol works even if there is more than one traitor. A nice aspect of the protocol is that if one adds one more round, in which $C$ distributes states for quantum communication from $A$ to $B$ and then from $B$ to $A$ (two rounds), then on similar grounds $C$ may know who of his subordinate generals is a traitor, if there is any. Thus one can have even more than it is required in Byzantine Agreement. Notice further, that the crux of the protocol is that the communication is kind of completely opposite to cryptography. It is totally inescapably insecure. And this is why the protocol works.

Thus, by now we have succeeded in establishing a secure quantum BA protocol for all possible values of $q$. According to Ref. \cite{RZS12} for $q=q_{max}$ the Hardy test (\ref{hardy0}) is fully device-independent. Hence, our BA protocol is also device-independent for $q=q_{max}$.

In summary, we have presented a device-independent quantum protocol for Byzantine Agreement, with any act of cheating leading to no transmission. It relies on an ability to establish a quantum link between three partners and perform pairwise Hardy tests between them. In correspondence to the original concept, the commanding officer C does not receive any information, as his task is only to distribute orders. He only accepts qubits, which carry no information. On the other hand, we put no restrictions on the amount of the exchange of auxiliary information. 

All the existing quantum protocols are only for detectable BA problem and none of them are for the original problem as they are described, since a quantum solution for such a problem is believed to be impossible. Here we have disproved this belief and presented a device-independent secure quantum scheme for the problem, as once the bits $m_{CA}$ and $m_{CB}$ are transmitted (readable to the receiver) then one must have $m_{CA}=m_{CB}$. { If correlations observed  between all pairs of parties satisfy Hardy's conditions (\ref{hardy0}) then} the protocol does not allow $m_{CA}\neq m_{CB}$, because any attempt by $C$ to do something like this results in {\it no transmission}, as the message is unreadable.

{\it Acknowledgments:} RR and MZ acknowledge support by Foundation for Polish Science TEAM project(TEAM/2011-8/9/styp7) co-financed by EU European Regional Development Fund and ERC grant QOLAPS(291348). MW acknowledges support from the Foundation for Polish Science (project HOMING PLUS/2011-4/14) and project QUASAR.

\end{document}